\documentclass[10pt]{article}
\usepackage{graphicx}
\usepackage{dcolumn}
\usepackage{bm}
\usepackage{color}
\usepackage{amssymb}
\usepackage{amsmath}
\usepackage{multicol,times,subfigure}

\begin{document}

\title{\textbf{A comparative analysis of the statistical properties of large mobile phone calling networks}}
\author{
Ming-Xia Li{\small$^{\mbox{1}}$},
Zhi-Qiang Jiang{\small$^{\mbox{1}}$},
Wen-Jie Xie{\small$^{\mbox{1}}$},
Salvatore Miccich\`{e}{\small$^{\mbox{2}}$},
Michele Tumminello{\small$^{\mbox{3}}$},\\
Wei-Xing Zhou{\small$^{\mbox{1}}$}
\& Rosario N. Mantegna{\small$^{\mbox{2,4}}$}
}

\maketitle

\vspace{-5mm}

\noindent
$^1$School of Business, School of Science, and Research Center for Econophysics, East China University of Science and Technology, Shanghai 200237, China,
$^2$Dipartimento di Fisica e Chimica, Universit\`a degli Studi di Palermo, Viale delle Scienze, Edificio 18, I-90128 Palermo, Italia,
$^3$Dipartimento di Scienze Economiche, Aziendali e Statistiche, Universit\`a degli Studi di Palermo, Viale delle Scienze, Edificio 13, I-90128 Palermo, Italia,
$^4$Center for Network Science and Department of Economics, Central European University, Nador utca 9, H-1051 Budapest, Hungary.

\medskip
\noindent
{\small{Correspondence and requests for materials should be addressed to W.X.Z. (wxzhou@ecust.edu.cn) or R.N.M (rn.mantegna@gmail.com).}}

\vspace{1mm}
\begin{center}
({\em{Received 25 February 2014; Accepted XXX; Published XXX}}) 
\end{center}
\vspace{1mm}

\begin{quote}
  {\noindent\textbf{Mobile phone calling is one of the most widely used communication methods in modern society. The records of calls among mobile phone users provide us a valuable proxy for the understanding of human communication patterns embedded in social networks. Mobile phone users call each other forming a directed calling network. If only reciprocal calls are considered, we obtain an undirected mutual calling network. The preferential communication behavior between two connected users can be statistically tested and it results in two Bonferroni networks with statistically validated edges. We perform a comparative analysis of the statistical properties of these four networks, which are constructed from the calling records of more than nine million individuals in Shanghai over a period of 110 days. We find that these networks share many common structural properties and also exhibit idiosyncratic features when compared with previously studied large mobile calling networks. The empirical findings provide us an intriguing picture of a representative large social network that might shed new lights on the modelling of large social networks.}}
\end{quote}


%

In the past two decades, the number of mobile phone users in China increased dramatically. There were 47.5 thousand users in 1991. This number increased to 84.5 million in 2000. In October 2013, it was released by the Ministry of Industry and Information Technology of China that there were 1.216 billion mobile phone users. The number of people using mobile phones is certainly less than that number, because it is not uncommon that a person owns two or more mobile phone numbers. Nevertheless, the actual population of mobile phone users is huge. Hence, the records of mobile phone users provide us great opportunities to study human's mobility patterns, communication dynamics, and social structure.

Gonz{\'a}lez {\it{et al.}} studied 16,264,308 displacements between successive mobile phone calls of 100,000 individuals randomly selected from a sample of more than 6 million anonymized mobile phone users over a six-month period in a European country and found that the density function follows a shifted power law with an exponential cutoff \cite{Gonzalez-Hidalgo-Barabasi-2008-Nature}. An analysis of human movements based on the trajectories of 464,670 dollar bills obtained from a bill-tracking system in the United States shows that jumps in human trajectories are distributed as a power law \cite{Brockmann-Hufnagel-Geisel-2006-Nature}. In contrast, there is evidence showing that intra-urban human mobility does not follow a power law but an exponential distribution according to mobile phone records \cite{Kang-Ma-Tong-Liu-2012-PA} and taxi trips data \cite{Peng-Jin-Wong-Shi-Lio-2012-PLoS1,Liang-Zheng-Lv-Zhu-Xu-2012-PA,Liang-Zhao-Dong-Xu-2013-SR}. It is crucial to point out that, when compared to human's mobility patterns at the aggregate level, individuals' patterns might not be homogeneous but exhibit different features \cite{Yan-Han-Wang-Zhou-2013-SR}. In addition, different data from different regions may also give different results \cite{Jiang-Yin-Zhao-2009-PRE}. Intriguingly, human's mobility patterns are largely predictable \cite{Song-Qu-Blumm-Barabasi-2010-Science,Lu-Bengtsson-Holme-2012-PNAS,Simini-Gonzalez-Maritan-Barabasi-2012-Nature}. It is unmistakable to foresee that mobile phone data will play a more significant role along the progress of constructing smart cities.

Understanding the temporal patterns of human's communication dynamics is essential in the tracking and management of information spreading and social contagion. According to the analysis of durations between two consecutive calls \cite{Gonzalez-Hidalgo-Barabasi-2008-Nature,Candia-Gonzalez-Wang-Schoenharl-Madey-Barabasi-2008-JPAMT,Karsai-Kaski-Kertesz-2012-PLoS1,Karsai-Kaski-Barabasi-Kertesz-2012-SR,Jiang-Xie-Li-Podobnik-Zhou-Stanley-2013-PNAS} and short-message correspondences \cite{Hong-Han-Zhou-Wang-2009-CPL,Wu-Zhou-Xiao-Kurths-Schellnhuber-2010-PNAS,Zhao-Xia-Shang-Zhou-2011-CPL}, it is well documented that the distribution of inter-communication durations has a fat tail and human interactions exhibit non-Poissonian characteristics. The non-Poissonian communication patterns are also observed in other situations such as email communications \cite{Barabasi-2005-Nature,Malmgren-Stouffer-Motter-Amaral-2008-PNAS} and letter correspondences \cite{Oliveira-Barabasi-2005-Nature,Li-Zhang-Zhou-2008-PA,Malmgren-Stouffer-Campanharo-Amaral-2009-Science}.

Mobile phone communication data also provide a useful channel for the study of social structure from the perspective of complex networks
\cite{Palla-Barabasi-Vicsek-2007-Nature,Onnela-Saramaki-Hyvonen-Szabo-Lazer-Kaski-Kertesz-Barabasi-2007-PNAS,Jo-Pan-Kaski-2011-PLoS1,Kumpula-Onnela-Saramaki-Kaski-Kertesz-2007-PRL}. For instance, we can infer friendship network structure by using mobile phone data, offering an alternative method to the standard self-report survey \cite{Eagle-Penland-Lazer-2009-PNAS}. The investigation of temporal motifs in the calling networks unveils the existence of temporal homophily in the sense that similar individuals tends to participate in mutual communications  \cite{Kovanen-Kaski-Kertesz-Saramaki-2013-PNAS}. The topological properties of a large calling network constructed from European data have been investigated in detail \cite{Onnela-Saramaki-Hyvonen-Szabo-deMenezes-Kaski-Barabasi-Kertesz-2007-NJP}, which enhanced our understanding of human social networks and shed new light on modelling weighted large-scale social networks.

In this paper, we investigate the statistical properties of four calling networks (directed calling network (DCN), mutual calling network (MCN), statistically validated directed calling network (SVDCN) and statistically validated mutual calling network (SVMCN), see {\textit{Methods}} section for the details of network construction) constructed from the detailed call records of more than nine million different mobile phone numbers from a mobile service provider in China. The DCN is a calling network, in which all mobile phone users in our data set are treated as nodes and a directed link is drawn from a call initiator to a call receiver. The MCN is a bidirectional calling network in which an edge is only drawn between any two mobile users with reciprocal calls. We also extract from DCN and MCN two Bonferroni networks SVDCN and SVMCN in which the links are filtered by a statistical validation test \cite{Tumminello-Micciche-Lillo-Piilo-Mantegna-2011-PLoS1}.


\section*{Results}

\noindent{\textbf{Size distribution of isolated components and the small-world effect.}}
Since we can only access the calling records of one mobile service provider, the constructed networks are fragmented into isolated subnetworks or ``components''. The original calling network (DCN) contains 236,738 components and its statistically validated calling network (SVDCN) has 468,138 components. There are 3,456,437 nodes and 16,269,689 edges in the giant component of the DCN (GCDCN) and 1,044,522 nodes and 1,440,366 edges for the giant component of the SVDCN, respectively. In contrast, there are 260,799 components in the mutual calling network (MCN) and 198,323 components in the statistically validated mutual calling network (SVMCN). The giant component of the MCN, denoted GCMCN, has 1,978,680 nodes and 4,677,642 edges, while the giant component of the SVMCN has 526,234 nodes and 765,213 edges. We summarize this information in Table \ref{Table:SumStat}.

\begin{table}[htb]
  \centering
  \caption{Sizes of the four calling networks and their giant components. $N_{\mathrm{node}}$ and $N_{\mathrm{edge}}$ are respectively the number of nodes and edges of a calling network.  $N_{\mathrm{Comp}}$ is the number of components of a calling network. $N_{\mathrm{GC,node}}$ and $N_{\mathrm{GC,edge}}$ are respectively the number of nodes and edges of the giant component of a calling network.}
  \label{Table:SumStat}
  \begin{tabular}{rrrrrrrrrrr}
  \hline\hline
     CN  &  $N_{\mathrm{node}}$ & $N_{\mathrm{edge}}$ & $N_{\mathrm{Comp}}$   & $N_{\mathrm{GC,node}}$  & $N_{\mathrm{GC,edge}}$ \\
  \hline
    DCN  &  4,032,884 & 16,753,635 & 236,738  &   3,456,437 & 16,269,689 \\
  SVDCN  &  2,410,757 &  2,453,678 & 468,138  &   1,044,522 &  1,440,366 \\
    MCN  &  2,615,247 &  5,065,397 & 260,799  &   1,978,680 &  4,677,642 \\
  SVMCN  &  1,042,751 &  1,099,254 & 198,323  &     526,234 &    765,213 \\
  \hline\hline
  \end{tabular}
\end{table}

\begin{figure}[htbp]
  \includegraphics[width=1\textwidth]{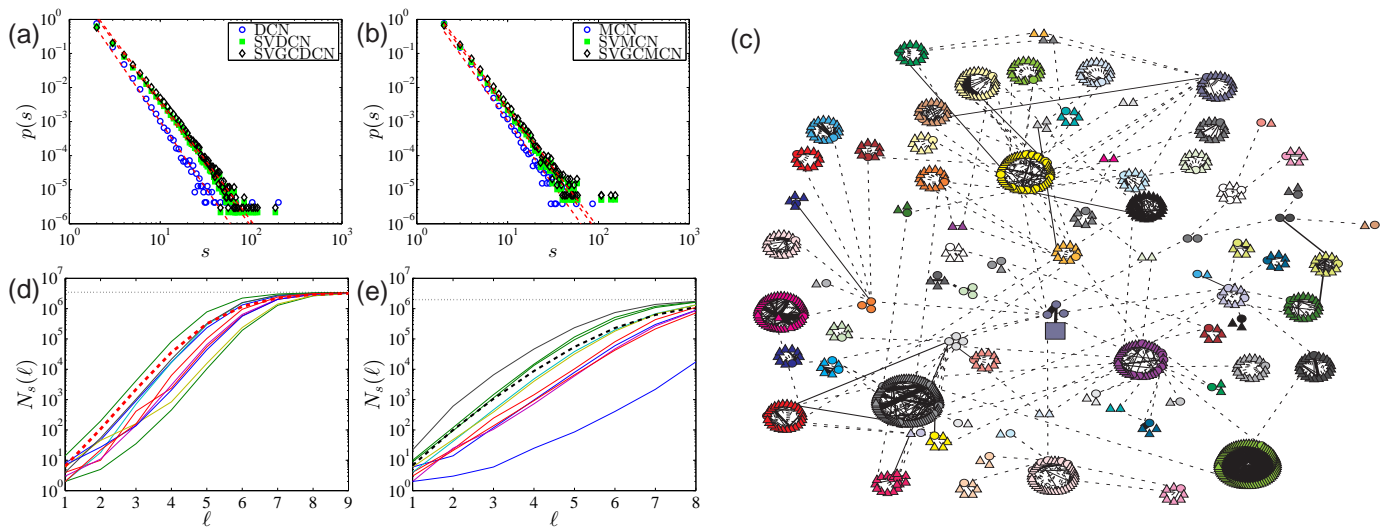}
  \vskip -0.5cm
  \caption{{\textbf{Network components.}} (a) Component size distributions of the calling network (DCN), the statistically validated calling network (SVDCN), and the statistically validated network of the DCN giant component (SVGCDCN). (b) Component size distributions of the mutual calling network (MCN), the statistically validated bidirectional calling network (SVMCN), and the statistically validated network of the MCN giant component (SVGCMCN). (c) An ego network extracted from the MCN, containing all nodes within a distance $\ell=5$ from the source node ($\square$) and the corresponding edges. The nodes having the maximum distance from the source node are drawn as triangles ($\triangle$) and other nodes are drawn as circles ($\circ$). The solid lines represent the validated calling relationship in the SVMCN, while the dashed lines are the original edges in the MCN. Nodes with the same color form a component. (d) Number $N_s(\ell)$ of nodes in the ego network within a distance of $\ell$ from the source node obtained by snowball sampling as a function of distance $\ell$ for the random choices of the source node (solid lines) and their average (dashed line) for the GCDCN. The dotted black line refers to the maximum size of the GCDCN. (e) Number $N_s(\ell)$ as a function of distance $\ell$ for the GCMCN.}
\label{Fig1:Components}
\end{figure}

Panels (a) and (b) of Fig.~\ref{Fig1:Components} show the empirical distributions of component size $s$, which is defined as the number of nodes in a component for the four communication networks. In Fig.~\ref{Fig1:Components}, the giant components are not included. It is found that the four distributions exhibit an asymptotic power-law decay
\begin{equation}
 p(s) \sim s^{-(1+\alpha)},
 \label{Eq:PDF:s}
\end{equation}
where the tail exponent $\alpha$ is $2.89$ for the DCN, $2.60$ for the SVDCN, $2.75$ for the MCN, and $2.58$ for the SVMCN. The statistical validated networks SVDCN and SVMCN have a shallower slope and therefore a wider distribution of component sizes than the DCN and MCN. This observation is due to the fact that the giant component of each original network has been segmented by removing the edges that are not statistical validated as illustrated in Fig.~\ref{Fig1:Components}(c). We also find that the component size distributions of the statistically validated networks of GCDCN and GCMCN have power-law tails and both tail exponents are $\alpha=2.54$.

We now turn to investigate the local structure of DCN and MCN through their ego networks \cite{Karsai-Kaski-Kertesz-2012-PLoS1}. For a randomly chosen source node, its ego network of distance $\ell$ contains all the nodes whose distance to the source node is no longer than $\ell$. An example of ego network extracted from the GCMCN is illustrated in Fig.~\ref{Fig1:Components}(c).The number of nodes of an ego network of distance $\ell$, $N_s(\ell)$, is plotted as a function of $\ell$ in Fig.~\ref{Fig1:Components}(d) and (e) for several random chosen source nodes and their average. It can be seen that the number of nodes increases exponentially when $\ell \leq 6$ and saturates to the size of the whole network with a slower growth rate when $\ell > 6$. Hence, the two giant components GCDCN and GCMCN exhibit a small-world effect \cite{Watts-Strogatz-1998-Nature}.

\begin{figure}[htbp]
  \centering
  \includegraphics[width=0.5\textwidth]{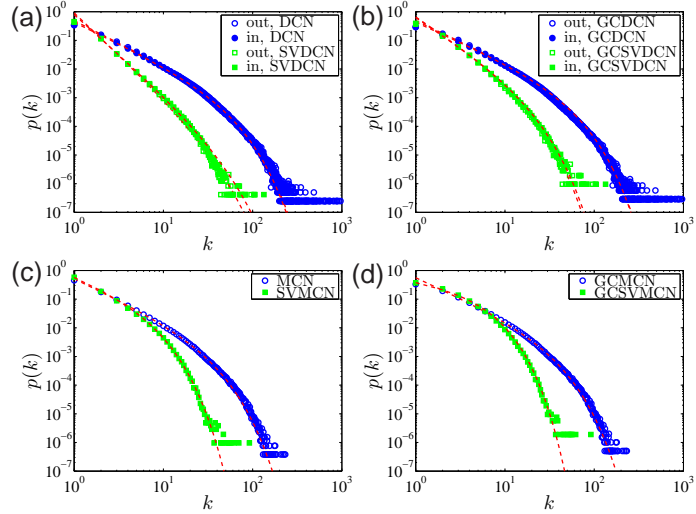}
  \vskip -0.5cm
  \caption{\textbf{Degree distribution.} (a) Distributions of in-degree and out-degree of the DCN and the SVDCN. (b) Distributions of in-degree and out-degree of the giant components of the DCN and SVDCN. (c) Degree distributions of the MCN and the SVMCN. (d) Degree distributions of the giant components of the MCN and the SVMCN. The dashed red lines are the fitted curves using exponentially truncated power law distributions.}
\label{Fig2:DegreePDF}
\end{figure}

\noindent{\textbf{Degree distribution.}}
Since the DCN and the SVDCN are directed, we investigate their in-degree and out-degree distributions as shown in Fig.~\ref{Fig2:DegreePDF}(a). All the four probability distributions can be well fitted by an exponentially truncated power law \cite{Clauset-Shalizi-Newman-2009-SIAMR}:
\begin{equation}
   p(k) = ak^{-\gamma_k}e^{-k / k_c},
   \label{Eq:DegreePDF}
\end{equation}
where $\gamma_k^{\mathrm{out}} = 1.52$ and $k_c^{\mathrm{out}} = 34.65$ for the out-degree distribution of the DCN, $\gamma_k^{\mathrm{in}} = 1.49$ and $k_c^{\mathrm{in}} = 40.36$ for the in-degree distribution of the DCN, $\gamma_k^{\mathrm{out}} = 2.90$ and $k_c^{\mathrm{out}} = 23.96$ for the out-degree distribution of the SVDCN, and $\gamma_k^{\mathrm{in}} = 2.76$ and $k_c^{\mathrm{in}} = 27.12$ for the in-degree distribution of the SVDCN. Figure~\ref{Fig2:DegreePDF}(b) plots the in-degree and out-degree distributions of the giant components of the DCN and the SVDCN, denoted GCDCN and GCSVDCN in the legend. These distributions can also be nicely fitted by the exponentially truncated power law of Eq.~(\ref{Eq:DegreePDF}). The estimated parameters are $\gamma_k^{\mathrm{out}} = 1.42$ and $k_c^{\mathrm{out}} = 34.60$ for the out-degree distribution of the GCDCN, $\gamma_k^{\mathrm{in}} = 1.38$ and $k_c^{\mathrm{in}} = 33.71$ for the in-degree distribution of the GCDCN, $\gamma_k^{\mathrm{out}} = 2.00$ and $k_c^{\mathrm{out}} = 10.00$ for the out-degree distribution of the GCSVDCN, and $\gamma_k^{\mathrm{in}} = 1.98$ and $k_c^{\mathrm{in}} = 10.37$ for the in-degree distribution of the GCSVDCN.

According to Fig.~\ref{Fig2:DegreePDF}(a) and (b), the corresponding distributions of a network and its giant component are very similar and share quite a few features. The first feature is that there is no evident difference between the in-degree and out-degree distributions for all the four networks. However, the distribution of an original network exhibits a much heavier tail than its statistically validated network. For instance, the average degree of the two giant components (GCMCN and GCSVMCN) are $\langle{k}^{\mathrm{GCMCN}}\rangle = 4.73$ and $\langle{k}^{\mathrm{GCSVMCN}}\rangle = 2.91$, which means that a mobile phone user on average reciprocally exchanges calls with more than 4 people of whom about 3 people are frequent contacts. However, there are outliers with very large in-degrees and out-degrees that cannot be modelled by the exponentially truncated power law. In addition, there are users characterized by a very large number of out-calls and a small or average number of in-calls. Most of these outliers are not typical mobile phone users but hot lines or ``robots'' \cite{Jiang-Xie-Li-Podobnik-Zhou-Stanley-2013-PNAS}. After filtering out the edges that do not pass the statistical validation, the number of outliers reduces significantly in the distributions of Bonferroni networks.

In Fig.~\ref{Fig2:DegreePDF}(c) and (d), we present the degree distributions of the MCN, of the SVMCN, and of the two giant components of these two networks (GCMCN and GCSVMCN). These four networks are not directed since the edges stand for reciprocal calls between any two users. These degree distributions can also be well fitted by the exponentially truncated power law of Eq.~(\ref{Eq:DegreePDF}). The estimated parameters are $\gamma_k = 1.46$ and $k_c = 20.81$ for the MCN, $\gamma_k = 1.20$ and $k_c = 4.27$ for the SVMCN, $\gamma_k = 1.40$ and $k_c = 21.00$ for the GCMCN, and $\gamma_k = 0.40$ and $k_c = 3.35$ for the GCSVMCN. For comparison, we note that the degree distribution for the European GCDCN has a shifted power-law form $p(k) = a (k +k_0)^{-\gamma_k}$ with $k_0 = 10.9$ and $\gamma_k = 8.4$ \cite{Onnela-Saramaki-Hyvonen-Szabo-Lazer-Kaski-Kertesz-Barabasi-2007-PNAS}. Most of the features of the MCN networks are similar to those of the DCN networks. An interesting difference is that the right-end tails become much narrower, because the reciprocal calling criterion for the construction of MCN has the ability to filter out most of those abnormal calls associated with hot lines and robots which are often unidirectional.

\noindent{\textbf{Degree-degree correlation.}}
The mixing patterns of complex networks have significant implications on the structure and function of the underlying complex systems \cite{Newman-2002-PRL}. Most social networks are reported to be assortative, i.e., people with many contacts are connected to others who also have many contacts. This may lead to a positive degree-degree correlation in the network, suggesting that the degree of a node positively correlates to the average degree of its neighborhood. The average nearest neighbors degree of a node $i$ is defined as $k_{nn,i} = (1/k_i) \sum_{j \in \mathcal{N}_i} k_j$, where $\mathcal{N}_i$ is the neighbor nodes set of $i$. In the calculation of $k_{nn}$ for the DCN and the SVDCN, we do not consider the direction of the edges. By averaging this value over all nodes in the network for a given degree $k$, one can calculate the average nearest neighbors degree denoted by $\langle k_{nn}|k\rangle$. A network is said to be assortatively mixed if $\langle k_{nn}|k\rangle$ increases with $k$ and disassortatively mixed if it decreases as a function of $k$.

\begin{figure}[htbp]
  \centering
  \includegraphics[width=0.5\textwidth]{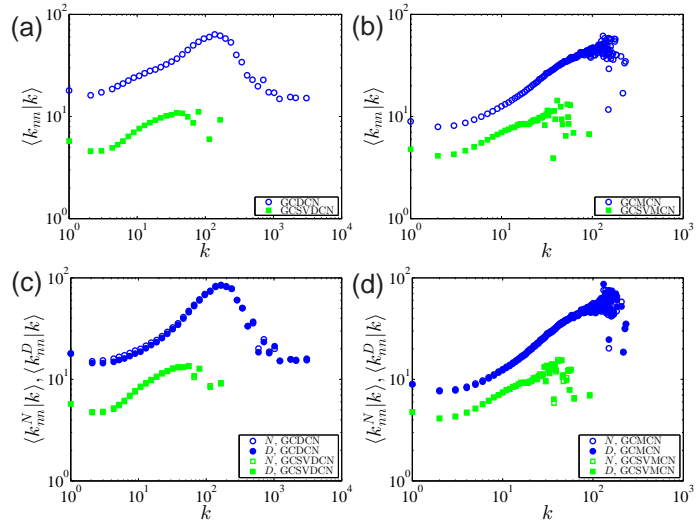}
  \vskip -0.5cm
  \caption{{\textbf{Degree-degree correlation.}} (a) Average nearest neighbor degree $\langle k_{nn}|k\rangle$ as a function of degree $k$ for the GCDCN and GCSVDCN. (b) Average nearest neighbor degree $\langle k_{nn}|k\rangle$ as a function of degree $k$ for the GCMCN and GCSVMCN. (c) Weighted average nearest neighbor degree $\langle k^N_{nn}|k\rangle$ and $\langle k^D_{nn}|k\rangle$ as a function of degree $k$ for the GCDCN and the GCSVDCN. (d) Weighted average nearest neighbor degree $\langle k^N_{nn}|k\rangle$ and $\langle k^D_{nn}|k\rangle$ as a function of degree $k$ for the GCMCN and the GCSVMCN.}
\label{Fig3:DegreeCorr}
\end{figure}

In Fig.~\ref{Fig3:DegreeCorr}(a) and (b), we show the dependence of $\langle k_{nn}|k\rangle$ as a function of $k$ for the giant components of the four networks. We find that $\langle k_{nn}|k=1\rangle>\langle k_{nn}|k=2\rangle$ for all curves. This observation was also present in the investigation of a large European dataset \cite{Onnela-Saramaki-Hyvonen-Szabo-deMenezes-Kaski-Barabasi-Kertesz-2007-NJP}. For $k$ values larger than 2, the $\langle k_{nn}|k\rangle$ function exhibits an evident increasing trend to reach a maximum. After the maximum, there is a decreasing region for large $k$. We notice that the overall shape of the two curves for the two MCN networks is qualitatively similar to that observed in the investigation of the European dataset \cite{Onnela-Saramaki-Hyvonen-Szabo-deMenezes-Kaski-Barabasi-Kertesz-2007-NJP}. A closer scrutiny of the GCDCN curve unveils an approximate plateau for the largest degrees. This can be partly explained by the fact that the nodes with the largest degrees usually correspond to hot lines or robots who receive calls from or call to diverse people. Figure~\ref{Fig3:DegreeCorr}(a) shows that mobile phone users with a ``reasonable'' number of contacts form an assortative network, while users with an abnormally large number of contacts exhibit a disassortative mixing pattern.

We also compute two weighted average nearest neighbors degrees defined as $k^N_{nn,i}=\sum_{j\in\mathcal{N}_i}k_jw^N_{ij}/s^N_i$ and $k^D_{nn,i}=\sum_{j\in\mathcal{N}_i}k_jw^D_{ij}/s^D_i$ to measure the degree-degree correlations \cite{Barrat-Barthelemy-PastorsSatorras-Vespignani-2004-PNAS,Onnela-Saramaki-Hyvonen-Szabo-deMenezes-Kaski-Barabasi-Kertesz-2007-NJP}, where $w^N_{ij}$ is the number of calls between $i$ and $j$, $s^N_i$ is the total number of calls between $i$ and her contacts, $w^D_{ij}$ is the call duration between $i$ and $j$, and $s^D_i$ is the total call duration between $i$ and her contacts. In the calculation of $k^N_{nn,i}$ and $k^D_{nn,i}$ for the DCN and the SVDCN, we do not consider the direction of the edges. We average these two weighted degrees over all users with the same degree $k$ to get $\langle k^N_{nn,i}|k\rangle$ and $\langle k^D_{nn,i}|k\rangle$. We show in Fig.~\ref{Fig3:DegreeCorr}(c) and (d) the weighted average nearest neighbour degrees for the four giant components of the four networks. We note that there is no significant difference between the two curves with number and duration weights for each network. The weight-based curves in Fig.~\ref{Fig3:DegreeCorr}(c) and (d) exhibit almost the same behaviors as the unweighted results in Fig.~\ref{Fig3:DegreeCorr}(a) and (b).

\begin{figure}[htbp]
\centering
  \includegraphics[width=0.5\textwidth]{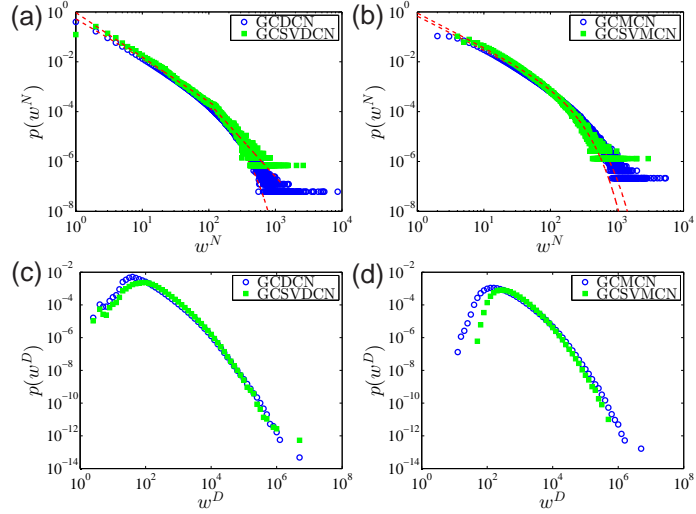}
  \vskip -0.5cm
  \caption{{\textbf{Edge weight distributions for the giant components of the four networks.}} (a) Distributions of number-based edge weight $w^N$ for the GCDCN and the GCSVDCN. (b)  Distributions of number-based edge weight $w^N$ for the GCMCN and the GCSVMCN. (c) Distributions of duration-based edge weight $w^D$ for the GCDCN and the GCSVDCN. (d)  Distributions of duration-based edge weight $w^D$ for the GCMCN and the GCSVMCN.}
\label{Fig4:EdgeWeightPDF}
\end{figure}

\noindent{\textbf{Edge weight distribution.}}
For a calling network, we have defined two kinds of weight for each edge between two users. For the DCN and the SVDCN, the number-based edge weight $w^N_{ij}$ is the number of calls occurred between user $i$ and user $j$ and the duration-based edge weight $w^D_{ij}$ is the total time elapsed during users $i$ and $i$ talk to each other through their mobile phones. Following Ref.~\cite{Onnela-Saramaki-Hyvonen-Szabo-deMenezes-Kaski-Barabasi-Kertesz-2007-NJP}, we focus on the giant components of the four networks. For the GCSVMCN, two connected users talked with each other on average $\langle w^N\rangle \thickapprox 23.98$ times and $\langle w^D\rangle \thickapprox 2234$ seconds ($37$ minutes). Figure \ref{Fig4:EdgeWeightPDF} shows the distributions of the giant components of the four networks.

Figure \ref{Fig4:EdgeWeightPDF}(a) shows that the distribution for the GCSVDCN exhibits an obvious kink at $w^N\approx120$. It is not clear why there is such a kink. We can use a bi-power-law distribution to fit the data
\begin{eqnarray}
   p(w) \sim w^{-\alpha_1}, & 1<w<120\\
   p(w) \sim w^{-\alpha_2}, & w>120
   \label{Eq:EdgeWeightPDF:GCSVDCN}
\end{eqnarray}
where the two power-law exponents are $\alpha_1 = 1.79$ and $\alpha_2 = 2.97$. In contrast, the distribution for the GCDCN can be fitted by an exponentially truncated power law
\begin{equation}
   p(w) = aw^{-\gamma_w}e^{-w / w_c},
   \label{Eq:EdgeWeightPDF}
\end{equation}
where $\gamma_w^N = 1.60$ and $w^N_c = 140.1$. The two distributions in Fig.~\ref{Fig4:EdgeWeightPDF}(b) can also be fitted by Eq.~(\ref{Eq:EdgeWeightPDF}), except for the region defined by $w^N<10$. The estimated parameters are $\gamma_w^N = 1.35$ and $w^N_c = 174.1$ for the GCMCN and $\gamma_w^N = 1.37$ and $w^N_c = 120.45$ for the GCSVMCN. We note that, rather than using the maximum likelihood estimation as in \cite{Clauset-Shalizi-Newman-2009-SIAMR}, the least-squares regression approach is adopted to fit the curves throughout this work. Indeed the method proposed in \cite{Clauset-Shalizi-Newman-2009-SIAMR} cannot be applied straightforwardly to truncated power-laws.

The distributions of duration-based edge weight $w^D$ for the giant component of the four networks are shown in Fig.~\ref{Fig4:EdgeWeightPDF}(c) and (d). The distributions for the original network and its corresponding statistically validated network are very similar. There is a maximum in each distribution occurring for a value, which is close to 100 seconds. These distributions cannot be well fitted by the exponentially truncated power law expressed in Eq.~(\ref{Eq:EdgeWeightPDF}), nor a power law.

\begin{figure}[htbp]
  \centering
  \includegraphics[width=0.5\textwidth]{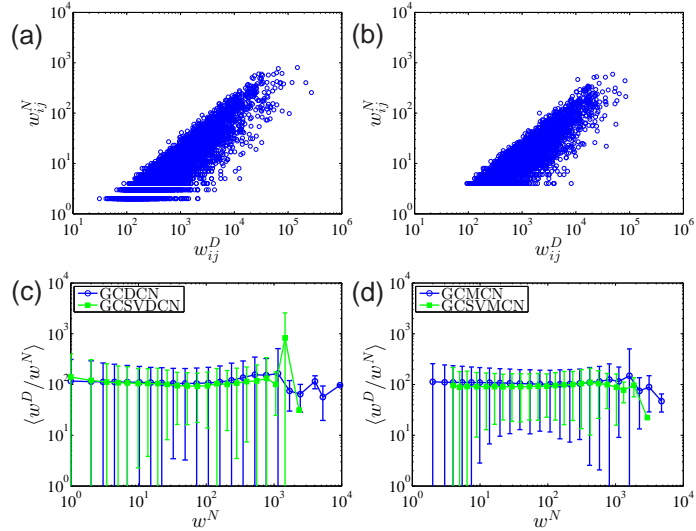}
  \vskip -0.5cm
  \caption{{\textbf{Edge weight correlations.}} (a) Scatter plot of duration-based weights $w^D_{ij}$ and number-based weights $w^N_{ij}$ of a random sample of 5000 edges in the giant component of the MCN. (b) Scatter plot of $w^D_{ij}$ and $w^N_{ij}$ of a random sample of 5000 edges in the giant component of the SVMCN. (c) Plot of $\langle w^D / w^N \rangle$ as a function of $w^N$ for the GCDCN and the GCSVDCN. (d) Plot of $\langle w^D / w^N \rangle$ as a function of $w^N$ for the GCMCN and the GCSVMCN.}
\label{Fig5:WeightCorr}
\end{figure}

\noindent{\textbf{Correlations between edge weights.}}
One would expect that there is a positive correlation between the weights of call number $w^N_{ij}$ and call duration $w^D_{ij}$. In Fig.~\ref{Fig5:WeightCorr}(a) and (b), we illustrate the scatter plots of duration-based weights $w^D_{ij}$ and number-based weights $w^N_{ij}$ of a random sample of 5000 edges selected in the giant component of the MCN and the SVMCN. The two weights are strongly correlated as expected. The Pearson's linear correlation coefficient $r$ between $w^N_{ij}$ and $w^D_{ij}$ is $r(w^N_{ij},w^D_{ij}) = 0.660$ for the GCMCN and $r(w^N_{ij},w^D_{ij}) = 0.726$ for the GCSVMCN, indicating the existence of a strong positive correlation. The relationship between the two link weights can also be characterized by Spearman's rank correlation coefficient $\rho$, which is a non-parametric measure of the statistical dependence between two variables. We obtain that $\rho(w^N_{ij},w^D_{ij}) = 0.8802$ for the GCMCN and $\rho(w^N_{ij},w^D_{ij}) = 0.864$ for the GCSVMCN. Since Spearman's correlation is higher than Pearson's correlation, the correlation has a nonlinear component in spite of the presence of a linear trend in the association between $w^N_{ij}$ and $w^D_{ij}$. The results for the GCDCN and the GCSVDCN are similar.

To analyze in more detail the correlation, we equally partition the interval $[\min(w^N), \max(w^N)]$ into 30 intervals by logarithmic binning and assign each link into one of the 30 groups. We obtain the call number weight $w^D$ as a function of call duration weight $w^N$ by calculating the mean and standard deviation of $w^N$ and $w^D$ in each group. Specifically, we plot $\langle w^D/w^N\rangle$ as a function of $w^N$ for the GCDCN and the GCSVDCN in Fig.~\ref{Fig5:WeightCorr}(c) and for the GCMCN and the GCSVMCN in Fig.~\ref{Fig5:WeightCorr}(d). The average duration of a call is close to 200 seconds for all the networks and it is almost independent of the number of calls. We observe that the statistical validated networks have lower $\langle w^D/w^N\rangle$ values and lower fluctuations. Another interesting feature is that the call duration fluctuates more for small or large number of calls.

\noindent{\textbf{Node strength distribution.}}
For each user, we define two node strengths based on the number and duration of calls. The number-based node strength $s^N_i = \sum_{j\in{\mathcal{N}}_i} w^N_{ij}$ is the total number of calls the user made, while the duration-based node strength $s^D_i = \sum_{j\in\mathcal{N}_i} w^D_{ij}$ is the total duration of calls the user performed. For the directed networks, we can further distinguish incoming and outgoing node strengths.

\begin{figure}[htbp]
  \centering
  \includegraphics[width=0.5\textwidth]{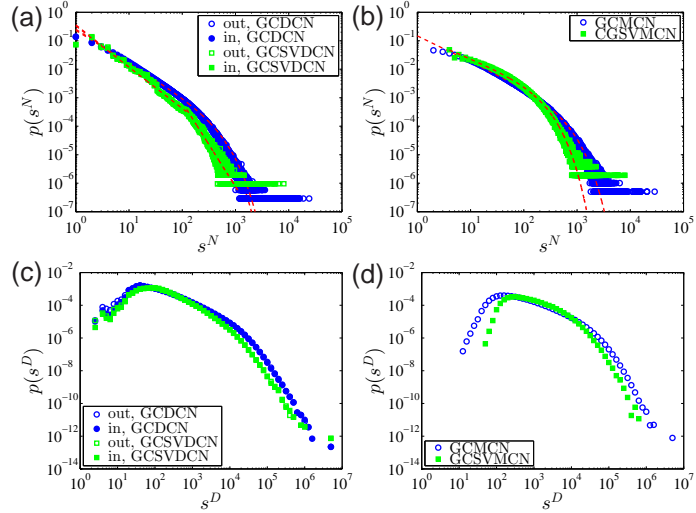}
  \vskip -0.5cm
  \caption{{\textbf{Node strength distributions.}} (a) Distributions of number-based node strength $s^N$ for the GCDCN and the GCSVDCN. (b) Distributions of number-based node strength $s^N$ for the GCMCN and the GCSVMCN. (c) Distributions of duration-based node strength $s^D$ for the GCDCN and the GCSVDCN. (d) Distributions of duration-based node strength $s^D$ for the GCMCN and the GCSVMCN.}
\label{Fig6:NodeStrengthPDF}
\end{figure}

The distributions of number-based node strength are shown in Fig.~\ref{Fig6:NodeStrengthPDF}(a) and (b) for the giant components of the four networks. The distributions for the GCDCN, the GCMCN and the GCSVMCN can be fitted by an exponentially truncated power law function
\begin{equation}
  p(s) \sim s^{-\gamma_s}\exp(-s / s_c).
  \label{Eq:NodeStrengthPDF}
\end{equation}
The fitted curves are shown as dashed red lines in Fig.~\ref{Fig6:NodeStrengthPDF}(a) and (b). We estimate that $\gamma_s^{N, {\mathrm{out}}} = 1.15$, $s^{N, {\mathrm{out}}}_c = 332.11$, $\gamma_s^{N, {\mathrm{in}}} = 1.15$ and $s^{N, {\mathrm{in}}}_c = 403.89$ for the GCDCN, $\gamma_{s}^N = 0.9$ and  $s^N_c = 470.5$ for the GCMCN, and $\gamma_s^N = 0.77$ and $s^N_c = 179$ for the GCSVMCN. For the GCSVDCN, the distribution curves are not smooth and the fitting would be of poor quality.

The distributions of duration-based node strength are shown in Fig.~\ref{Fig6:NodeStrengthPDF}(c) and (d) for the giant components of the four networks. These distributions share a very similar shape, which is reminiscent of the inter-call durations at the population level \cite{Candia-Gonzalez-Wang-Schoenharl-Madey-Barabasi-2008-JPAMT,Karsai-Kaski-Barabasi-Kertesz-2012-SR,Jiang-Xie-Li-Podobnik-Zhou-Stanley-2013-PNAS}. For the directed networks, there is no difference between incoming and outgoing call durations. Figure~\ref{Fig6:NodeStrengthPDF}(d) shows that the statistical validation method is able to filter out the nodes with very short or very long mutual call durations.

\begin{figure}[htbp]
  \centering
  \includegraphics[width=0.5\textwidth]{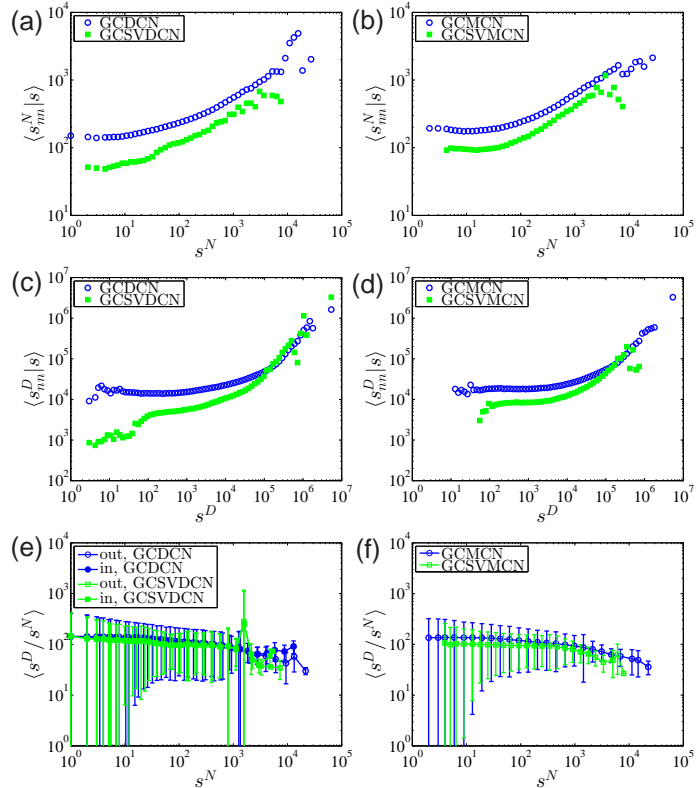}
  \vskip -0.5cm
  \caption{{\textbf{Node strength correlations.}} (a) Average number-based node strength $\langle s^N_{nn} | s^N \rangle$ as a function of $s^N$ for the GCDCN and the GCSVDCN. (b) Average number-based node strength $\langle s^N_{nn} | s^N \rangle$ as a function of $s^N$ for the GCMCN and the GCSVMCN. (c) Average duration-based node strength $\langle s^D_{nn} | s^D \rangle$ as a function of $s^D$ for the GCDCN and the GCSVDCN. (d) Average duration-based node strength $\langle s^D_{nn} | s^D \rangle$ as a function of $s^D$ for the GCMCN and the GCSVMCN. (e) Plot of $\langle s^D / s^N \rangle$ as a function of $s^N$ for the GCDCN and the GCSVDCN. (f) Plot of $\langle s^D / s^N \rangle$ as a function of $s^N$ for the GCMCN and the GCSVMCN.}
\label{Fig7:NodeStrengthCorr}
\end{figure}

\noindent{\textbf{Correlations between node strength.}}
For nodes, besides the degree-degree correlation, we also study the correlation between node strength. The number-based and duration-based correlation of node strengths are calculated as follows: $s^N_{nn} = (1 / k_i) \sum_{j \in \mathcal{N}_i} s^N_j$ and $s^D_{nn} = (1 / k_i) \sum_{j \in \mathcal{N}_i} s^D_j$. The results for the giant components of the four networks are illustrated in Fig.~\ref{Fig7:NodeStrengthCorr}(a)-(d). In the figure, all curves show a very slow increase for small $s^D$ and $s^N$ values and then a more pronounced increase for large values of $s$. For small $s < s_x$, independence can be observed, whereas a dependence of the form $\langle s_{nn}|s\rangle \sim s^{\alpha}$  is observed for large $s^D$ and $s^N$ values. For strong ties with $w^D > 2 \times 10^4$, which form $1.6\%$ of all edges of the original giant component, the strength of both adjacent nodes depends almost on the weight of this single edge ($s_i = w_{ij} = s_j$). This explains the linear trend in the strength-strength correlation of the original GC network ($\alpha^{D,{\mathrm{MCN}}} \approx 1$). In contrast, we find that $\alpha^{N,\mathrm{MCN}} \approx 0.5$ when $s^{N,\mathrm{MCN}} > 200$, $\alpha^{D,\mathrm{MCN}} \approx 1$ when $s^{D,\mathrm{MCN}} > 10^4$, $\alpha^{N,\mathrm{SVMCN}} \approx 0.5$ when $s^{N,\mathrm{SVMCN}} > 50$, and $\alpha^{D,\mathrm{SVMCN}} \approx 0.67$ when $s^{D,\mathrm{SVMCN}} > 200$.

Similar to Fig.~\ref{Fig5:WeightCorr}(c) and (d), we calculate and plot $\langle s^D/s^N\rangle$ as a function of $s^N$ for the giant components in Fig.~\ref{Fig7:NodeStrengthCorr}(e) and (f). It is found that all the curves exhibit a slight decreasing trend both in the mean and in the standard deviation as a function of node strength. In addition, the curves for the statistically validated networks are lower than their original counterparts.

\begin{figure}[htbp]
  \centering
  \includegraphics[width=1\textwidth]{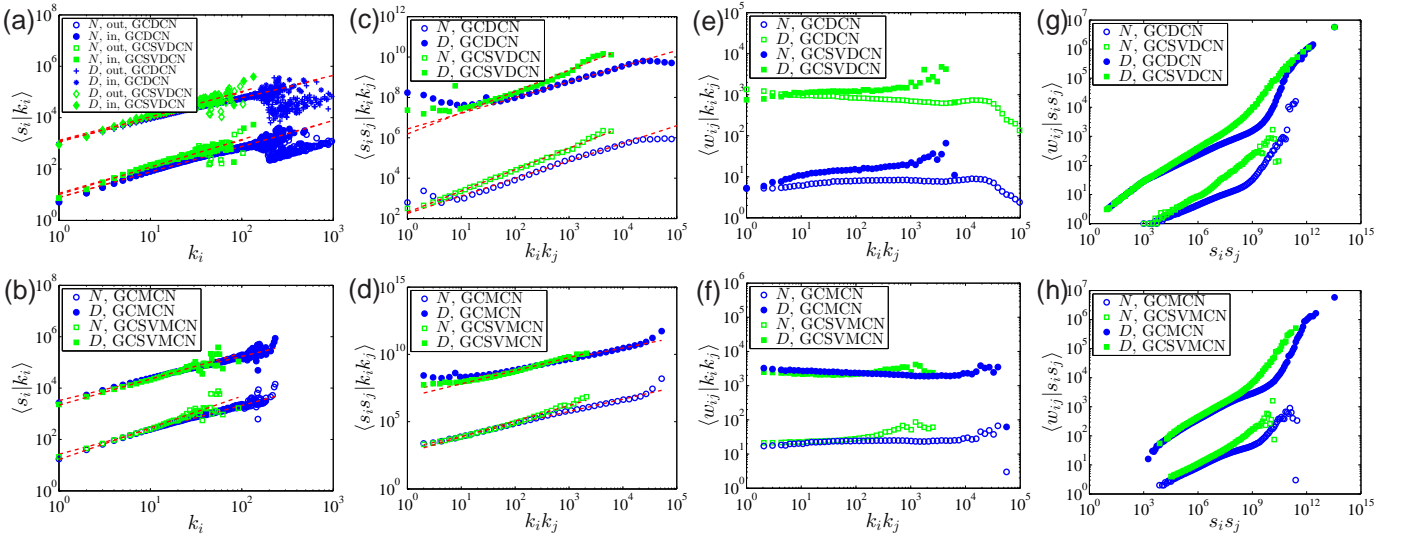}
  \vskip -0.5cm
  \caption{{\textbf{Cross-correlations between node strength, edge weight and node degree.}} (a,b) Power-law dependence of the average number-based and duration-based node strength on the node degree for the giant components of the four networks. (c,d) Dependence of $\langle s^N_is^N_j | k_ik_j \rangle$ and $\langle s^D_is^D_j | k_ik_j \rangle$ on the degree product. (e,f) Average duration-based edge weight $\langle w^D_{ij} | k_ik_j \rangle$ and number-based edge weight $\langle w^N_{ij} | k_ik_j \rangle$ as a function of degree product $k_ik_j$. (g,h) Average duration-based edge weight $\langle w^D_{ij} | s^D_is^D_j \rangle$ and number-based edge weight $\langle w^N_{ij} | s^N_is^N_j \rangle$ as a function of strength product $s^D_is^D_j$. The curves for number-weighted node strength have been shifted rightwards horizontally by a factor of 1000 for clarity.}
\label{Fig8:xCorr:w:s:k}
\end{figure}

\noindent{\textbf{Cross-correlations between node strength, edge weight and node degree.}}
We now turn to the cross-correlations between node strength, edge weight and node degree. Figure~\ref{Fig8:xCorr:w:s:k}(a) and (b) plot the dependence of the node strengths on the node degree for the giant components of the four networks. The curves have a power law dependence: $\langle s|k\rangle \sim k^{\alpha}$. The best fitting power-law exponents are the following: $\alpha^{N,{\mathrm{out}}} = \alpha^{N,{\mathrm{in}}} \approx 1.0$ and $\alpha^{D,{\mathrm{out}}} = \alpha^{D,{\mathrm{in}}} \approx 0.85$ for the GCDCN, $\alpha^{N,{\mathrm{out}}} = \alpha^{N,{\mathrm{in}}}\approx 1.1$ and $\alpha^{D,{\mathrm{out}}} = \alpha^{D,{\mathrm{out}}} \approx 0.95$ for the GCSVDCN, $\alpha^{N} \approx 0.95$ and $\alpha^{D} \approx 0.86$ for the GCMCN, and $\alpha^{N} \approx 1.01$ and $\alpha^{D} \approx 1.23$ for the GCSVMCN. For the GCMCN, the average call durations of individuals who have high degrees are slightly less than that of individuals with low degrees. These results confirm that the statistical validation procedure filters out communications occurring between users linked by underlying social relationships. We also observe that $\alpha^N > \alpha^D$ for the GCMCN, suggesting that individuals who talk to a large quantity of users appear to spend less time per callee than those who spend less time on phone.

We present the correlation between strength product $s_is_j$ and degree product $k_ik_j$ in Fig.~\ref{Fig8:xCorr:w:s:k}(c) and (d). Also in this case we observe a clear power-law dependence $\langle s_is_j | k_ik_j \rangle \sim \langle k_ik_j \rangle^\beta$. According to Ref.~\cite{Onnela-Saramaki-Hyvonen-Szabo-deMenezes-Kaski-Barabasi-Kertesz-2007-NJP}, if $\langle s_i \rangle = k_i \langle w \rangle$, one would expect that $\langle s_is_j | k_ik_j \rangle = \langle w \rangle^2 \langle k_ik_j \rangle$. Differently than expected, we obtain that $\beta^{N} \approx 1.12$ and $\beta^{D} \approx 1$ for the GCDCN, $\beta^{N} \approx 1.35$ and $\beta^{D} \approx 1.48$ for the GCSVDCN, $\beta^{N} \approx 0.9$, $\beta^{D} \approx 0.8$ for the GCMCN, and $\beta^{N} \approx 1.2$ and $\beta^{D} \approx 1$ for the GCSVMCN. The discrepancy of $\beta\neq1$ indicates that there are correlations between node degree and the weights of the edges adjacent to the node.

We also study the correlation between edge weight and node degree product (Fig.~\ref{Fig8:xCorr:w:s:k}(e) and (f)) and the correlation between edge weight and node strength product (Fig.~\ref{Fig8:xCorr:w:s:k}(g) and (h)). The $\langle w^D_{ij} | k_ik_j \rangle$ curve and the $\langle w^N_{ij} | k_ik_j \rangle$ curve are very similar for each network, and there are evident difference between the $\langle w_{ij} | k_ik_j \rangle$ curves of an original network and its statistically validated network. However, the dependence of the $\langle w_{ij} | k_ik_j \rangle$ curves on the degree product $k_ik_j$ is weak. In contrast, the $\langle w_{ij} | s^N_is^N_j \rangle$ curves increase rapidly and exhibit roughly power laws: $\langle w_{ij}|s_is_j\rangle \sim (s_is_j)^\delta$, where $\delta^{N}\approx0.43$ and $\delta^{D}\approx0.44$ for the GCDCN, $\delta^{N}\approx0.42$ and $\delta^{D}\approx0.47$ for the GCSVDCN, $\delta^{N} \approx 0.3$ and $\delta^{D}\approx0.45$ for the GCMCN, and $\delta^{N} \approx 0.4$ and $\delta^{D} \approx 0.5$ for the GCSVMCN.

\noindent{\textbf{Clustering coefficients.}}
Clustering coefficient is an important metric of complex networks. It represents the local cohesiveness around a node. The clustering coefficient of node $i$ is defined as $C_i = 2 t_i / [k_i(k_i - 1)]$, where $t_i$ is the number of triangles of node $i$ with its neighbours. For the directed networks (DCN and SVDCN), we treat edges as undirected. We find that the average clustering coefficients of the giant components of the four networks are 0.11 (DCN), 0.02 (SVDCN), 0.12 (MCN), and 0.11 (SVMCN). The relatively small values of the average clustering coefficients suggest that tree-shaped subgraphs are quite frequent in the local structure of the four networks. Indeed, the clustering coefficient of about 72.5\% of the users is zero. It is worth  noting that  the clustering coefficients of the communication networks of European users are also small \cite{Onnela-Saramaki-Hyvonen-Szabo-deMenezes-Kaski-Barabasi-Kertesz-2007-NJP}. We also observe that the average clustering coefficient in the SVDCN is smaller than in the DCN network. This observation reflects the fact that the statistical validation approach, while minimizing the presence of links not related to an underlying social relationship, may also remove edges with meaningful social relationships. See the Methods section for a more detailed discussion of this aspect.

\begin{figure}[htbp]
  \centering
  \includegraphics[width=0.5\textwidth]{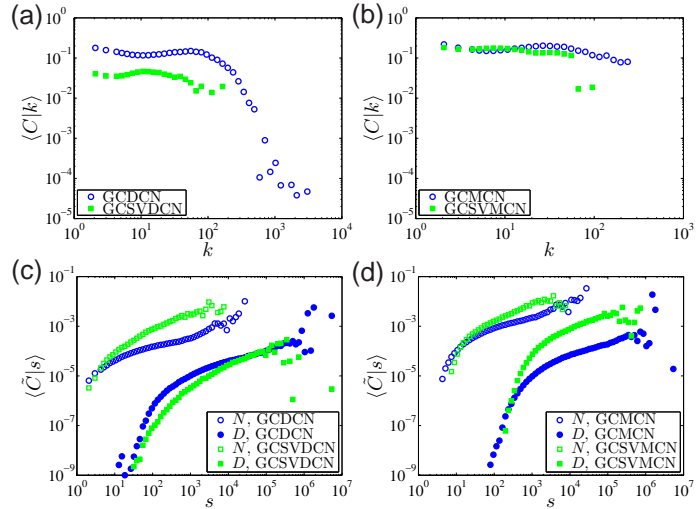}
  \vskip -0.5cm
  \caption{{\textbf{Clustering coefficient.}} (a) Average clustering coefficient $\langle C |k \rangle$ as a function of $k$ for the GCDCN and the GCSVDCN. (b) Average clustering coefficient $\langle C |k \rangle$ as a function of $k$ for the GCMCN and the GCSVMCN. (c) Average weighted clustering coefficient $\langle\tilde{C} | s^N\rangle$ and $\langle\tilde{C} | s^D\rangle$ as a function of $s$ for the GCDCN and the GCSVDCN. (d) Average weighted clustering coefficient $\langle\tilde{C} | s^N\rangle$ and $\langle\tilde{C} | s^D\rangle$ as a function of $s$ for the GCMCN and the GCSVMCN.}
\label{Fig9:ClusteringCoeff}
\end{figure}

Panels (a) and (b) of Fig.~\ref{Fig9:ClusteringCoeff} show the dependence of $\langle{C|k}\rangle$ on $k$ for the four networks. Surprisingly, we do not observe a power-law decay as observed for the European users \cite{Onnela-Saramaki-Hyvonen-Szabo-deMenezes-Kaski-Barabasi-Kertesz-2007-NJP}. On the contrary, high-degree users have a relatively high clustering coefficient. This can be partially explained by the fact that one main promotion strategy of  the mobile phone service provide is to make contract with institutions with lower communication prices. The users with more contacts are usually ``secretaries'' and their contacts also call each other frequently. Figure~\ref{Fig9:ClusteringCoeff}(c) and (d) present the dependence of average weighted clustering coefficient $\langle{\tilde{C}|s}\rangle$ \cite{Onnela-Saramaki-Hyvonen-Szabo-deMenezes-Kaski-Barabasi-Kertesz-2007-NJP} on $s$ for the four networks. The increasing trend in these curves is also observed in the European case \cite{Onnela-Saramaki-Hyvonen-Szabo-deMenezes-Kaski-Barabasi-Kertesz-2007-NJP}.

\noindent{\textbf{Topological overlap of two connected nodes.}}
The topological overlap of the neighborhood of two connected nodes $i$ and $j$ is estimated by considering the relative overlap of their common neighbors \cite{Onnela-Saramaki-Hyvonen-Szabo-deMenezes-Kaski-Barabasi-Kertesz-2007-NJP},
\begin{equation}
\label{EQ:Overlap}
O_{ij} = \frac{n_{ij}}{k_i + k_j - 2 - n_{ij}}
\end{equation}
where $k_i$ and $k_j$ are the degrees of the two nodes and $n_{ij}$ is the number of neighbors common to both nodes $i$ and $j$. Overlap is the fraction of common neighbors that a pair of connected nodes has, which is different from the edges-clustering coefficient reflecting the probability that a pair of connected vertices has a common neighbor \cite{Radicchi-Castellano-Cecconi-Loreto-Parisi-2004-PNAS}. In the calculation of overlap for the directed networks, we ignore the direction of edges and treat the directed networks as undirected networks.

\begin{figure}[htbp]
  \centering
  \includegraphics[width=0.5\textwidth]{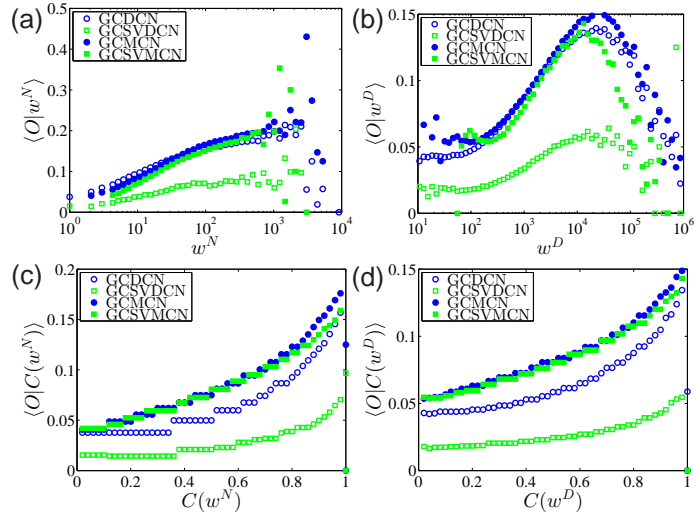}
  \vskip -0.5cm
  \caption{{\textbf{Topological overlap.}} (a) Average overlap $\langle O | w^N \rangle$ as a function of number-based edge weight $w^N$ for the four networks. (b) Average overlap $\langle O | w^D \rangle$ as a function of duration-based edge weight $w^D$ for the four networks. (c) Average overlap $\langle O | C(w^N) \rangle$ as a function of cumulative number-based edge weight $w^N$ for the four networks. (d) Average overlap $\langle O | C(w^D) \rangle$ as a function of cumulative duration-based edge weight $C(w^D)$ for the four networks. }
\label{Fig10:Overlap:w}
\end{figure}

Fig.~\ref{Fig10:Overlap:w}(a) illustrates the average overlap $\langle O | w^N \rangle$ as a function of the number-based edge weight $w^N$ for the four networks. The two curves for the MCN and the SVMCN are similar, while the curve for the DCN is higher than that for the SVDCN indicating that a significant fraction of common neighbors have been removed by the statistical validation method. In addition, all the curves exhibit an increasing trend and the two blue curves seemingly decrease after $w^N\approx2000$. The curve for the MCN is very different from the European case with a bell shape curve with a maximum at $w^N\approx50$ \cite{Onnela-Saramaki-Hyvonen-Szabo-deMenezes-Kaski-Barabasi-Kertesz-2007-NJP}. Fig.~\ref{Fig10:Overlap:w}(b) illustrates the average overlap $\langle O | w^D \rangle$ as a function of duration-based edge weight $w^D$ for the four networks. Similar to the European case \cite{Onnela-Saramaki-Hyvonen-Szabo-deMenezes-Kaski-Barabasi-Kertesz-2007-NJP}, all the average overlap curves $\langle O | w^D \rangle$ increase up to $w^D \approx 2\times10^4$, whereas after that they decline quickly. In Fig.~\ref{Fig10:Overlap:w}(c) and (d), we show the average overlap $\langle O | P_c(w^N) \rangle$ and $\langle O | P_c(w^D) \rangle$ against the cumulative edge weight $P_c(w^N)$ and $P_c(w^D)$ respectively. Different from the behavior observed in the European case \cite{Onnela-Saramaki-Hyvonen-Szabo-deMenezes-Kaski-Barabasi-Kertesz-2007-NJP}, all the curves increase. Fig.~\ref{Fig10:Overlap:w} shows that the statistical validation method does not change much the overlap structure of the mutual calling networks. However, the overlap reduces remarkably after applying the statistical validation method on the edges of the directed calling networks.

\section*{Discussion}

We have constructed and investigated four calling networks from a data set of more than nine million phone users. These networks are the directed calling network, the mutual calling network and their statistically validated networks. The statistical properties of these four calling networks have been investigated in a comparative way. Specifically, we have considered the distributions of the degree, the edge weight, the node strength, the relative overlap of two connected nodes, and their mutual dependence. We found that these networks share many common topological properties and also exhibit idiosyncratic characteristics in both qualitative and quantitative ways. When compared with the results observed for a mutual calling network of an European data set of mobile phone users \cite{Onnela-Saramaki-Hyvonen-Szabo-deMenezes-Kaski-Barabasi-Kertesz-2007-NJP}, the results obtained for the Shanghai data set exhibit some different communication behaviors.

The differences between the two original calling networks (DCN and MCN) and their statistically validated networks are of great interest. We have observed that the size of statistically validated network is significantly smaller than its original network. Also, the Bonferroni networks have thinner degree distributions, indicating fewer highly connected nodes. This finding suggests that a large proportion of edges in high-degree nodes might not be directly associated with an underlying social motivation, which is consistent with the finding that there are hot lines and robots calling a large number of different users and characterized by an ultra low number of incoming calls \cite{Jiang-Xie-Li-Podobnik-Zhou-Stanley-2013-PNAS}. For the original networks the average call durations of high-degree users are slightly less than that of low-degree users, while for the statistical validated networks, we observe the opposite situation that high-degree individuals have larger average call durations than low-degree individuals. Our comparative analysis shows the importance of investigating statistically validated networks because the original networks contain users whose communication patterns are not reflecting a social motivation. The calling profile of these users makes difficult to uncover the true communication behavior of the system.

It is natural that the networks for the Shanghai data set and the European data set share many common topological properties. However, we also observed several discrepancies. The differences are of crucial interest as they might point to different mechanisms at play in mobile communication networks (and more generally in social networks) located at different parts of the world. For instance, we observed a different broadness of degree distributions, which might originate from different dynamics in social ties formation and disappearance \cite{Ghoshal-Chi-Barabasi-2013-SR}. The different behaviors that might explain formation and deletion of social ties indicate the presence of different elementary mechanisms governing social dynamics under different cultures and social norms. However, a detailed investigation of these issues is beyond the scope of this work.

The setting of the statistical validation and its threshold depends on the problem investigated. One can choose to use a more or less conservative threshold (as it is done when one choose a 0.05 or 0.01 or 0.001 univariate threshold). To investigate the possible impacts of different thresholds, we repeat all the analyses by using as a Bonferroni threshold $0.01/N_E$, where $N_E$ is the number of pairs of subscribers that had at least one call over the entire period for the DCN or the number of pairs of subscribers with mutual calls in the MCN. In this way we have two new Bonferroni networks for the DCN and MCN networks obtained with the least restrictive Bonferroni threshold we can set. It is obvious that the new Bonferroni networks have larger sizes. We find that the results are qualitatively the same as the more restrictive Bonferroni threshold. The differences are only quantitative. For instance, the degree distributions are broader simply because the there are more nodes with higher degrees.

It might be worth discussing more in detail the implications of the statistical validation. While the statistical validation is useful to filter edges like hot lines and robots, it also removes a consistent fraction of possible edges with meaningful social relationships. We argue that any other filtering methods also suffer a similar shortcoming. For instance the filtering method keeping bidirectional links while removing unidirectional links \cite{Onnela-Saramaki-Hyvonen-Szabo-deMenezes-Kaski-Barabasi-Kertesz-2007-NJP} or the method extracting the ``multiscale backbone'' of the original network in which the edges are statistically validated by identifying which links of a node carry disproportionate fraction of the weights \cite{Serrano-Boguna-Vespignani-2009-PNAS}. For such large social networks, we will never be able to identify all the {\emph{true}} social ties but rather any filtering procedure will present false positive links, that is, links present in the filtered network but without a social origin, and false negative links, that is, links that are absent in the filtered network but have a social origin. In this respect, we can say that our statistical validation method minimize the number of false positive links while does not put constraints on the number of false negative links. For example, a similar approach has been pursued $(i)$ in Ref. \cite{Hatzopoulos-Iori-Mantegna-Micciche-Tumminello-2014-QF} to investigate preferential credit links between banks and firms based on their mutual credit relationships or $(ii)$ in Ref. \cite{Tumminello-Lillo-Piilo-Mantegna-2012-NJP} to  identify clusters of investors from their real trading activity in a financial market. Further details about the methodology, specifically applied to mobile phone data, can also be found in Ref. \cite{Li-Palchykov-Jiang-Kaski-Kertesz-Micciche-Tumminello-Zhou-Mantegna-2014-NJP}, where interesting evolution patterns of triadic motifs are discussed.

\section*{Methods}

\noindent{\textbf{Data description.}}
Our data set comprises the detailed call records of more than nine million different mobile phone numbers from one mobile operator in Shanghai during two separated periods. One is from 28 June 2010 to 24 July 2010 and the other is from 1 October 2010 to 31 December 2010. Because the records in several hours are missing on October 12, November 6, 21, 27, and December 6, 8, 21, 22, these days are excluded from our sample. The sample has a total of 110 days of call records. Each entry of the records contains the following information, caller number, callee number, call starting time, call length, as well as call status. The caller and callee numbers are encrypted for protecting personal privacy. Call status with a value of 1 means that the call gets through successfully and is terminated normally. When we construct communication networks, only the calls with the call status equaling to 1 are considered.

\noindent{\textbf{Construction of networks.}}
There are three mobile operators in mainland China. We only have access to the entire call records used for billing purpose of one operator. We thus focus on the calling networks between mobile phone users that are costumers of the operator. We construct four calling networks as follows.

The directed calling network (DCN) is composed of all users. If user $i$ calls user $j$, a directed edge is assigned from $i$ to $j$. There are a total of 4,032,884 nodes and 16,753,635 directed edges in the DCN. The mutual calling network (MCN) contains part of the users. An edge is drawn between user $i$ and user $j$ if and only if there are reciprocal calls between them. All isolate nodes are not included in the MCN. There are totally 2,615,247 nodes and 5,065,397 edges in the MCN. One can see that about 70\% edges are not reciprocal in the DCN.

We then perform statistical validation on each edge of the DCN and the MCN as described below. Edges that are consistent with the null hypothesis of random selection of the receiver are removed together with the nodes that become isolated. With our procedure we obtain a statistical validated directed calling network (SVDCN) which has 2,410,757 nodes and 2,453,678 edges and a statistical validated mutual calling network (SVMCN) which has 1,042,751 nodes and 1,099,254 edges. The sizes of the two original networks reduce significantly.

\noindent{\textbf{Statistical validation of edges.}}
The statistical validation is performed by comparing the number of calls observed between each pair of caller and receiver with a null hypothesis of random matching between the caller and the receiver. The null hypothesis takes into account the heterogeneity in the number of calls performed by subscribers. The method is a variant of the approach originally proposed in Ref.  \cite{Tumminello-Micciche-Lillo-Piilo-Mantegna-2011-PLoS1} and used in different systems \cite{Tumminello-Lillo-Piilo-Mantegna-2012-NJP,Hatzopoulos-Iori-Mantegna-Micciche-Tumminello-2014-QF,Li-Palchykov-Jiang-Kaski-Kertesz-Micciche-Tumminello-Zhou-Mantegna-2014-NJP}. Here the statistical validation is done by considering the number of calls done by a caller, the number of calls received by a receiver and checking whether or not the number of calls exchanged between them is compatible with the null hypothesis that these calls were made by selecting randomly the receiver.  The test is performed as detailed hereafter. The test allows to assign a $p$-value to each pair of caller and  receiver. The $p$-values are then compared with a statistical threshold set at 1\%. However, since the null hypothesis of random pairing is tested for all pairs of subscribers, we have to perform a multiple hypothesis test correction in order to control the number of false positives. In this work, we use the Bonferroni correction which is the most restrictive amongst all possible corrections minimizing the number of false positives. When a link between two subscribers $i$ and $j$ has a $p$-value less than the Bonferroni threshold we assume that the calls from $i$ to $j$ have a social origin.

The $p$-value is obtained as follows. Let us denote $N$ as the total number of phone calls of all users in the calling network, $N_{ic}$ the number of calls initiated by individual $i$ and $N_{jr}$ the number of calls received by individual $j$. Assuming that $X = N_{icjr}$ is number of calls initiated by individual $i$ and received by individual $j$. The probability of observing $X$ co-occurrences is given as follows \cite{Tumminello-Micciche-Lillo-Piilo-Mantegna-2011-PLoS1,Tumminello-Micciche-Lillo-Varho-Piilo-Mantegna-2011-JSM}:
\begin{equation}
  \label{EQ:OverExpression}
  H(X|N,N_{ic},N_{jr}) = \frac{C^X_{N_{ic}}C^{N_{jr} - X}_{N - N_{ic}}}{C^{N_{jr}}_N},
\end{equation}
where $C^X_{N_{ic}}$ is a binomial coefficient. We can associate a $p$-value to the observed $N_{icjr}$ as follows:
\begin{equation}
  \label{EQ:p:Nicjr}
  p(N_{icjr}) = 1 - \sum^{N_{icjr} - 1}_{X = 0} H(X|N,N_{ic},N_{jr}).
\end{equation}

The Bonferroni correction for the multiple testing hypothesis is $p_b = 0.01/N_T$ where $N_T$ is the number of performed tests. For the DCN, we perform $N_T=16,753,635$ tests. If the estimated $p(N_{icjr})$ is less than $p_b$, we conclude that the calls between user $i$ and user $j$ cannot be explained by a null hypothesis of random calls from $i$ to $j$ performed according to the heterogeneity of the caller and the receiver. When the test does not reject the null hypothesis, the directed edge from $i$ to $j$ is removed.

In the validation of the MCN network, we need to estimate the $p$-value of the number of calls $N_{jcir}$ initiated by $j$ and received by $i$ in a similar way. For the MCN, we need to conduct $N_T=2\times5,065,397=10,130,794$ tests. The Bonferroni correction for the multiple hypothesis test is again $p_b = 0.01/N_T$. If the estimated $p(N_{icjr})$ is less than $p_b$, we can conclude that $i$ preferentially calls $j$. We also need to estimate the $p$-value of the number of calls $N_{jcir}$ initiated by $j$ and received by $i$ in a similar way. The edge between $i$ and $j$ is included in the statistically validated network if and only if the two directional links are both validated.

To illustrate how this method works, we present a few quantitative examples with typical values of calls extracted from Fig.~\ref{Fig1:Components}(c). We consider the DCN, in which $p_b=0.01/N_T=1/16,753,635=9.87\times10^{-10}$. The root node (square) is linked to one node close to it. For calls the root node made, $N_{ic}=14$, $N_{jr}= 81$, and $N_{icjr}=14$, leading to $p(N_{icjr})=6.36\times10^{-88}$. For calls the root node received, $N_{ic}=58$, $N_{jr}= 81$, and $N_{icjr}=14$, resulting in $p(N_{icjr})=0$. These two directed links between the root node and her unique contact are thus statistically significant. Consider the dashed link connecting a node in the lime green cluster and a node in the gray cluster, located in the southeast of Fig.~\ref{Fig1:Components}(c). For the direct link from the lime green node to the gray node, $N_{ic}=400$, $N_{jr}= 824$, $N_{icjr}=2$, and $p(N_{icjr})=3.41\times10^{-6}$. For the direct link from the lime green node to the gray node, $N_{ic}=289$, $N_{jr}= 459$, $N_{icjr}=1$, and $p(N_{icjr})=1.05\times10^{-3}$. In spite of the low $p$-values, these two directed links are statistically compatible with the null hypothesis of random selection of the receiver when a Bonferroni correction is applied. More information about the distribution of $p$-values can be found in Ref.~\cite{Li-Palchykov-Jiang-Kaski-Kertesz-Micciche-Tumminello-Zhou-Mantegna-2014-NJP}.

It is worth pointing out that many of the links not statistically validated might also be associated with a social origin. In fact, with our choice of the Bonferroni correction we primarily control the absence of false positives. This is done at the cost of observing an admittedly high level of false negative. The motivation behind our choice is that we aim to detect with our methodology a backbone of social interaction that is not affected by the presence of false positives.

%

\vspace{-5mm}
\section*{Acknowledgements}
\vspace{-3mm}
\noindent{This work was partially supported by the National Natural Science Foundation of China (11205057), the Humanities and Social Sciences Fund of the Ministry of Education of China (09YJCZH040), the Fok Ying Tong Education Foundation (132013), and the Fundamental Research Funds for the Central Universities.}

\vspace{-5mm}
\section*{Author contributions}
\vspace{-3mm}
\noindent{WXZ and RNM conceived the study. MXL, ZQJ, WJX, SM, MT, WXZ and RNM designed and performed the research. MXL performed the statistical analysis of the data. ZQJ, WXZ and RNM wrote the manuscript. MXL, ZQJ, WJX, SM, MT, WXZ and RNM reviewed and approved the manuscript.}

\vspace{-5mm}
\section*{Additional information}
\vspace{-3mm}


\noindent{{\textbf{Competing financial interests:}} The authors declare no competing financial interests.}

\vspace{2mm}

\noindent{{\textbf{License:}} This work is licensed under a Creative Commons Attribution-NonCommercial-NoDerivative Works 3.0 Unported License. To view a copy
of this license, visit http://creativecommons.org/licenses/by-nc-nd/3.0/}

\vspace{2mm}

\noindent{\textbf{How to cite this article:} Li, M.-X., Jiang, Z.-Q., Xie, W.-J., Miccich\`{e}, S., Tumminello, M., Zhou, W.-X. \& Mantegna, R. N. A comparative analysis of the statistical properties of large mobile phone calling networks. Sci. Rep. 4, xxx; DOI:10.1038/srep00xxx (2014).}

\newpage
\clearpage

\end{document}